\documentclass[prd,eqsecnum,aps,twocolumn, noshowpacs]{revtex4}
\usepackage{epsf,graphics,graphicx}
\usepackage{amsmath}
\usepackage{amssymb,latexsym,mathrsfs}
\usepackage{hyperref}
\usepackage{color}

\def\bea{\begin{eqnarray}}
\def\eea{\end{eqnarray}}
\def\ba{\begin{array}}
\def\ea{\end{array}}

\def\beq{\begin{equation}}
\def\eeq{\end{equation}}

\begin{document}

\title{The Berry phase in inflationary cosmology}

\author{Barun Kumar Pal\footnote{Electronic address: {barunp1985@rediffmail.com}}, ${}^{}$ Supratik Pal\footnote{Electronic address: {supratik@isical.ac.in}}
and  B. Basu$^{1}$\footnote{Electronic address:
{banasri@isical.ac.in}} ${}^{}$}
\affiliation{$^1$Physics and Applied Mathematics Unit, Indian Statistical Institute, 203 B.T. Road, Kolkata 700 108, India}

\vspace{1in}

\begin{abstract}
We derive an analogue of the Berry phase associated with inflationary
cosmological perturbations of quantum mechanical origin by obtaining the corresponding
wavefunction. We have further shown that cosmological Berry phase can be
completely envisioned through the observable parameters, viz. spectral indices. 
Finally, physical significance of this phase is discussed from the point of view of theoretical and observational
aspects with some possible consequences of this quantity in inflationary cosmology.
\end{abstract}


\maketitle

\section{Introduction}
Since its inception Berry phase has drawn a lot of attention in
physics community. Although the geometric phase was known long ago
\textit{a la} Aharonov-Bohm effect, the general context of a
quantum-mechanical state developing adiabatically in time under a
slowly varying parameter-dependent Hamiltonian  has been analyzed by
Berry \cite{berry} who argued that when the parameters return
adiabatically to their initial values after traversing a closed
path, the wavefunction acquires a {\it geometric phase factor}
depending on the path, in addition to the well-known {\it dynamical
phase factor}. The existence of a geometric phase in an adiabatic evolution
is not only confined to quantum phenomenon, the classical analogue of it also exists
and is referred to as the Hannay angle \cite{hannay}. Berry established
a semi-classical relation between the quantum and classical geometric phases in adiabatic evolution \cite{berry1}.
The Berry phase has been the subject of a variety of
theoretical and experimental investigations \cite{book}; possible
applications range from quantum optics and molecular physics to
fundamental quantum mechanics and quantum computation. Some analyses
have been made to study this phase in the area of cosmology and
gravitation also. In particular, the Berry phase has been calculated
in the context of relic gravitons \cite{furtado}. In \cite{cai} a
covariant generalization of the Berry phase has been obtained.
Investigations were also made to study the behavior of a scalar
particle in a class of stationary spacetime backgrounds and the
emergence of Berry phases in the dynamics of a particle in the
presence of a rotating cosmic string \cite{corichi}. The
gravitational analogue of the Aharonov-Bohm effect in the spinning
cosmic string spacetime background was also obtained \cite{mazur}.
Within a typical framework of cosmological model  the Berry  phase
has been shown to be associated with the decay width of the state in
case of some well known examples of vacuum instability \cite{dutta}.

The inflationary scenario \cite{guth} -- so far the most physically
motivated  paradigm for  early universe -- is also in vogue for
quite some time now. Among other motivations, the inflationary
scenario is successful to a great extent in explaining the origin of
cosmological perturbation seeds \cite{pert}. The accelerated
expansion converts the initial vacuum quantum fluctuations into
macroscopic cosmological perturbations. So, measurement of any quantum property which reflects on
classical observables serve as a supplementary probe of inflationary cosmology, complementing the well-known
 CMB polarization measurements \cite{wmap, planck}. Though this is an important issue, we notice that there has
been very little study in the literature which deals with proposing measurable quantities which may measure the
genuine quantum property of the seeds of classical cosmological perturbations. The only proposition that has drawn
our attention is via violation of Bell's inequality \cite{campo}. This has led us to investigate for the potentiality
 of Berry phase in  providing a  measurable quantum property which is inherent in the macroscopic character of classical
cosmological perturbations.

The paradigm of inflation provides a well motivated mechanism
for the origin of fluctuations observed in the large scale structure of the matter and in CMB.
In order to quantify these fluctuations the theory of general relativistic cosmological
perturbations has to be employed. However, in practice the field equations cannot be solved with its
full generality, some approximation has to be used. The linear order approximation to the cosmological perturbation
theory has been developed to a high degree of sophistication during the last few decades \cite{bardeen,mukhanov, mukhanov1}.
Although attempts have been made to develop
the second order perturbation theory, which provide results with better accuracy.  There are discussions on the
deviation from the first order approximation from the observation \cite{komatsu} 
and theoretical approaches
\cite{bartolo} through the non-Gaussianity, non-adiabaticity and so on. There are works which focus \cite{nakamura}
on the development
of the second order gauge invariant cosmological perturbation technique. Attention has also been paid \cite{losic} to
investigate the importance of second order corrections to linearized cosmological perturbation
theory in the inflationary scenario, which suggests that for many parameters of slow roll inflation the second order
contributions may dominate over the first order effects during super Hubble evolution.
All these recent efforts indicate the importance of the study of  second order cosmological perturbations,
but at a first go linear order perturbation theory has remained as the primary tool to investigate the behavior of
fluctuations during inflation. The main motivation behind this development was to clarify the relation
between the scenarios of the early universe and cosmological data, such as the CMB anisotropies. The developments in the observations were also supported by the
theoretical sophistication of the linear order cosmological perturbation theory.
Recently, the first order approximation of our universe from a homogeneous isotropic one
has also been  revealed through the observation of the CMB by the
Wilkinson Microwave Anisotropy Probe (WMAP) \cite{wmap}. 
At this juncture we also remind that our objective here is to see whether a link can be established between the
quantum property of the seeds of classical cosmological perturbations and
inflationary cosmological parameters through the derivation of associated Berry phase, for which consideration of
the sub Hubble modes are appropriate.  
Keeping all these points under consideration, we concentrate on the evolution of  the sub Hubble perturbation modes
within the framework of linearized perturbation theory.

In this article our primary purpose is to demonstrate the effect of the curved spacetime background in the dynamical evolution of the  quantum fluctuations during inflation through the derivation of the associated Berry phase and search for the possible consequences via observable parameters. 
The quantum fluctuations in inflaton are realized by Mukhanov-Sasaki
equation which is analogous to time dependent harmonic oscillator equation. The associated physical mechanism for cosmological perturbations can be reduced to the quantization of a parametric oscillator leading to particle creation due to interaction with the gravitational field and may be termed as cosmological Schwinger effect \cite{martin}. The relation between the Berry phase and the Hannay angle has been studied for the generalized time dependent harmonic oscillator \cite{ghoshdutta}. This relation is also extended \cite{child} from adiabatic to non-adiabatic time dependent harmonic oscillator. Stimulated by these, one may expect to derive the cosmological analogue of Berry phase in the context of inflationary perturbations
and search for possible consequences via observable parameters. To this end, we first find an exact wavefunction for the system of inflationary cosmological perturbation by solving the associated {\it Schroedinger equation}. The relation \cite{morales,monteoliva} between the dynamical invariant \cite{lewis1,lewis2,lewis3} and the geometric phase has then been utilized to derive the corresponding Berry phase.
For slow roll inflation the total accumulated phase gained by each of the modes during sub-Hubble oscillations
(adiabatic limit) is found to be a new parameter made of corresponding (scalar and tensor) spectral indices.
So in principle, measurement of the Berry phase of the quantum cosmological perturbations provide us an indirect route to estimate spectral
indices and other observable parameters therefrom. Further, since tensor spectral index is related to the tensor to
scalar amplitude ratio through the consistency relation, the Berry phase can indeed be utilized to act as a supplementary probe of inflationary cosmology.

\section{Linear Cosmological perturbation as a time dependent harmonic oscillator }
In cosmology, inhomogeneities grow because of the attractive nature of gravity. So inhomogeneities were smaller
in the past. Since we are interested in inflationary cosmological perturbations, inhomogeneities can be treated as linear
perturbations around the homogeneous and isotropic FLRW universe, the metric of which can be written as
\beq\label{frw}
ds^2=a^2(\eta)\left[-d\eta^2+\delta_{ij}dx^idx^j\right]
\eeq

During inflation the energy density of the universe was dominated by the potential energy, $V$,
of scalar field(s). Therefore the total action of the system can be cast into the form
\beq\label{gr}
S=\frac{M_P^2}{2}\int d^4x\sqrt{-g}R-\int d^4\sqrt{-g}\left[\frac{1}{2}g^{\mu\nu}\partial_\mu\phi
\partial_\nu\phi+V(\phi)\right]
\eeq
where $\phi$ is the scalar field driving inflation known as {\it inflaton field}. Also the most general linear perturbation, without the vector modes,
 about the homogeneous metric can
be expressed as \cite{bardeen}
\bea\label{pmetric}
\hspace{-.5cm}ds^2&=&a^2(\eta)\left[-d\eta^2(1-2A)+2\partial_iBdx^id\eta\right.\nonumber\\
&+&\left.\left([1-\psi]\delta_{ij}+2\partial_i\partial_jE+h_{ij}\right)dx^idx^j\right]
\eea
where the functions $A, ~B,~\psi$ and $E$ represent the scalar sector whereas the tensor $h_{ij}$, satisfying $h_i^i
=0=(\delta^{jm}\partial_m)h_{ij}$, represents the gravitational waves. Here we are interested in cosmological
perturbations induced by a single scalar field, as a consequence there will not be any vector perturbations. At the
linear level, scalar and tensor perturbations decouple and can therefore be studied separately.


The scalar fluctuations of geometry can be characterized by a single quantity, the gauge-invariant Bardeen potential
\cite{bardeen} defined by
\beq\label{bardeen}
\Phi_B\equiv A+\frac{1}{a}\frac{d}{d\eta}\left[a\left(B-\frac{dE}{d\eta}\right)\right]
\eeq
The fluctuations in the inflaton field are characterized by the following gauge-invariant quantity
\beq
\delta\phi^{GI}\equiv\delta\phi+\frac{d\phi}{d\eta}\left(B-\frac{dE}{d\eta}\right)
\eeq
These two gauge invariant quantities are coupled through the perturbed Einstein equations, and
in the scalar sector everything can be reduced to the study of a single gauge-invariant variable defined by
\cite{mukhanov}
\beq\label{ms}
v\equiv a\left[\delta\phi^{GI}+\frac{d\phi}{d\eta}\frac{\Phi_B}{\cal H}\right]
\eeq
Here the variable $v$ is related to the comoving curvature perturbation ${\cal
R}$ via the relation $v=-z{\cal R}$ where $z\equiv
\frac{a\phi^{'}}{{\cal H}(\eta)}$ and ${\cal
H}\equiv\frac{a^{'}}{a}$ being the  conformal Hubble
parameter and $'$ denotes derivative w.r.t. conformal time $\eta$.
The action for the scalar perturbation only can then be written as
\beq\label{sa}
{\textbf{ \cal S}}^S=\frac{1}{2}\int d\eta d\textbf{x}\left[v'^2-\delta^{ij}\partial_iv\partial_jv+\frac{z''}{z}v^2\right]
\eeq
To arrive at the action \eqref{sa}, one needs to expand the action \eqref{gr} upto the second order in the metric
perturbations and the scalar field fluctuations and have to use the background Einstein equations.

We should make a point of note here that we are using flat slicing. To reduce the action for the curvature perturbation
to its simplest form Eqn.\eqref{sa} in terms of a single gauge invariant variable $v$ utilizes the constraint equations obtained by varying
the action with respect to the first order perturbation variables \cite{mukhanov1}.
The complete expression for the second order action is found by taking into account the background equations
and constraint equation obtained by varying the action with respect to $B-E'$ has the form
\beq\label{tsa}
{\textbf{ \cal S}}^S=\frac{1}{2}\int d\eta d\textbf{x}\left[v'^2-\delta^{ij}\partial_iv\partial_jv+\frac{z''}{z}v^2
+\frac{M_P^2}{2}\mathcal{D}\right]\nonumber
\eeq
where $\mathcal{D}$ is in the form of a total divergence containing quadratic combination of first order perturbation
variables as a result we can drop this term from the action. 
In case of spatially closed slicing one could get the
same action as \eqref{sa} but then $v$ would have been different \cite{mukhanov1}. The same analysis can also be carried out
 in a Hamiltonian framework as has been done in \cite{lang1994}.


The Hamiltonian corresponding to the action \eqref{sa} is
\beq
{\bf H}=\frac{1}{2}\int d^3x\left[\Pi^2+\delta^{ij}\partial_iv\partial_jv-\frac{z''}{z}v^2\right]
\eeq
where $\Pi= v'$.

Now promoting the fields to operators and taking the following Fourier decompositions
\bea\label{ft}
\hat{v}({\bf x},\eta)&=&\int \frac{d^3k}{(\sqrt{2\pi})^3}\hat{v}_k e^{i{\bf k}.{\bf x}}\nonumber\\
\hat{\Pi}({\bf x},\eta)&=&\int\frac{d^3k}{(\sqrt{2\pi})^3} \hat{\Pi}_ke^{i{\bf k}.{\bf x}}
\eea
we found the Hamiltonian density operator corresponding to the above action \eqref{sa} to be
\bea
\hspace{-0.5cm}\hat{\textbf{H}}_{\bf
k}^S&=&\frac{1}{2}\left[\hat{\Pi}_{1\bf
k}^2+\left(k^2-\frac{z''}{z}\right)\hat{v}_{1\bf k}^2\right]+\frac{1}{2}\left[\hat{\Pi}_{2\bf
k}^2+\left(k^2-\frac{z''}{z}\right)\hat{v}_{2\bf k}^2\right]\nonumber\\
&\equiv& \hat{\textbf{H}}_{1\bf k}^S+\hat{\textbf{H}}_{2\bf k}^S
\eea
where we have
decomposed $\hat{v}_{\bf k}\equiv{\hat{v}}_{1\bf k}+i~{\hat{v}}_{2\bf
k}$ and $\hat{\Pi}_{\bf k}\equiv\hat{\Pi}_{1\bf k}+i~\hat{\Pi}_{2\bf
k}$ into their real and imaginary parts.

Similarly, considering the tensor perturbation only we have the corresponding action
\beq\label{ta}
{\cal \textbf{S}}^T=\frac{M_P^2}{2}\int d\eta d{\bf
x}\frac{a^2}{2}\left[h^{'2}-\delta^{ij}\partial_ih\partial_jh\right]
\eeq
By means of the substitution $u=\frac{M_P}{\sqrt{2}}{h}{a}$, promoting the fields to operators and taking the Fourier decomposition,
the Hamiltonian operator corresponding to the above action \eqref{ta} turns out to be
\bea
\hspace{-0.5cm}\hat{\textbf{H}}_{\bf
k}^T&=&\frac{1}{2}\left[\hat{\pi}_{1\bf
k}^2+\left(k^2-\frac{a''}{a}\right)\hat{u}_{1\bf k}^2\right]+\frac{1}{2}\left[\hat{\pi}_{2\bf
k}^2+\left(k^2-\frac{a''}{a}\right)\hat{u}_{2\bf k}^2\right]\nonumber\\
&\equiv& \hat{\textbf{H}}_{1\bf k}^T+\hat{\textbf{H}}_{2\bf k}^T
\eea
where we have decomposed $\hat{u}_{\bf k}\equiv {\hat{u}}_{1\bf k}+i~{\hat{u}}_{2\bf k}$ and
 $\hat{\pi}_{\bf k}\equiv{\hat{\pi}}_{1\bf k}+i~{\hat{\pi}}_{2\bf
k}$ similarly.

Thus, for both the scalar and tensor modes, the Hamiltonians are sum of
two time dependent harmonic oscillators, each of
them having the following form
\beq\label{subh}
\hat{\textbf{H}}_{j\bf
k}=\frac{1}{2}\bigg[\hat{p}_{j\bf k}^2+\omega^2\hat{q}_{j\bf
k}^2\bigg]
\eeq
where $\hat{q}_{j\bf k}=\hat{v}_{j\bf k}$, $\hat{u}_{j\bf k}$;
$\hat{p}_{j\bf k}=\hat{\Pi}_{j\bf k}$, $\hat{\pi}_{j\bf k}$  and
$\omega=\sqrt{k^2-\frac{z''}{z}}$, $\sqrt{k^2-\frac{a''}{a}}$ for scalars and tensors respectively.
One may note here that for the complete solution of the Schroedinger equation for the Hamiltonian (\ref{subh})
we have to deal with two situations: \\
(i) $k^2> \frac{z''}{z},\frac{a''}{a}$  where $\omega$ is real and corresponds to the sub Hubble modes,\\
(ii) $k^2< \frac{z''}{z},\frac{a''}{a}$ corresponds to the super Hubble modes with imaginary $\omega$.
 For the later case, the Hamiltonian can be re-written as
\beq\label{suph}
\hat{\textbf{H}}_{j\bf
k}=\frac{1}{2}\bigg[\hat{p}_{j\bf k}^2-\omega_I^2\hat{q}_{j\bf
k}^2\bigg]
\eeq
which represents an inverted harmonic oscillator
with time dependent frequency given by $i\omega_I=i\sqrt{\frac{z''}{z}-k^2}$, $ i\sqrt{\frac{a''}{a}-k^2}$
for scalars and tensors respectively.


 Though in further derivations we are not concerned with the super Hubble modes, but at this stage we are ready to provide the solutions for the whole spectrum.

We  find the solution to the Schroedinger equation for the Hamiltonians \eqref{subh} and \eqref{suph} using
{\it dynamical invariant operator method} \cite{lewis3} i.e. the Lewis-Risenfeld invariant formulation which has now become evident that it
 can be applied to the treatment of time dependent quantum system if a
invariant can be found.  To be precise, we want to analyze the situation by solving the associated
Schroedinger equation
\begin{equation}\label{ham1}
{\hat{\textbf{H}}}_{\bf k}{\Psi}=({\hat {\textbf{H}}}_{1{\bf k}}+~{\hat{\textbf{H}}}_{2{\bf
k}})\Psi=i\frac{\partial }{\partial \eta}\Psi
\end{equation}

In this {\it invariant} method, we first look for a nontrivial hermitian operator
$I_k(\eta)$ satisfying the Liouville-von Neumann equation
\begin{equation}\label{inv}
\frac{d I_k}{d\eta}=-i \left[I_k\textbf{,H}_k\right]+\frac{\partial
I_k}{\partial \eta}=0.
\end{equation}
Whenever such an invariant operator exists provided it does not
contain time derivative operator, one can write down the
solutions of the Schroedinger equation in the following form
\beq\label{soln}
\Psi_n=e^{i\alpha_n(\eta)}\Theta_n,~~n=0,1,2...
\eeq
 where $\Theta_n$ are the eigenfunctions of the operator $I_{\bf k}$ and $\alpha_n(\eta)$ are known as the Lewis phase.
Here,  $\textbf{H}_k=\textbf{H}_{1\bf k}+\textbf{H}_{2\bf k}$ and the invariant operator
associated to this Hamiltonian can be expressed as
\begin{equation}\label{invk}
I_k(\eta)\equiv I_1(q_{1{\bf k}},\eta)+I_{2}(q_{2{\bf k}},\eta)
\end{equation}

In the following we shall first derive these invariant operators and will find the solution to the Schroedinger
equation for two different cases.
\subsection{Modes with $k^2> \frac{z''}{z},\frac{a''}{a}$}
The modes in this regime, known as sub Hubble modes, oscillate with
real frequencies and the corresponding Hamiltonian has the form \eqref{subh}.
Following the usual technique \cite{lewis1,lewis2,lewis3, ghosh} we
obtain
\bea\label{invariant}
 I_k&=&\frac{1}{2}\left[
 \frac{q_{1{\bf k}}^2}{\rho_k^2}+\bigg(\rho_k p_{1{\bf k}}-\rho_k^\prime
 q_{1{\bf k}}\bigg)^2\right]\nonumber\\
 &+&\frac{1}{2}\left[
 \frac{q_{2{\bf k}}^2}{\rho_k^2}+\bigg(\rho_k p_{2{\bf k}}-\rho_k^\prime
q_{2{\bf k}}\bigg)^2\right]\nonumber\\
 &=& I_{1}+I_{2}
 \eea
where $\rho_k$ is a time dependent real function satisfying the
following Milne-Pinney equation 
\begin{equation}\label{mp}
{\rho_k}^{\prime\prime}+
\omega^2(\eta,k)\rho_k=\frac{1}{\rho^3_k(\eta)}
\end{equation}
To find the solutions of the Schroedinger
Eqn.(\ref{ham1}) we also need to know the eigenstates of the operator $I_k$ governed by the eigenvalue equation
\bea\label{evet}
I_{\bf k}\Theta_{n_1,n_2}\left(q_{1{\bf k}},q_{2{\bf
k}},\eta\right)=\lambda_{n_1,n_2}\Theta_{n_1,n_2}\left(q_{1{\bf
k}},q_{2{\bf k}},\eta\right)
\eea
The eigenstates of the operator $I_k$ turns out to be \cite{furtado}
\bea\label{tes}
\Theta_{n_1,n_2}&=&\frac{\bar{H}_{n_1}\left[\frac{q_{1{\bf
k}}}{\rho_k}\right]\bar{H}_{n_2}\left[\frac{q_{2{\bf
k}}}{\rho_k}\right]}{\sqrt[4]{\pi^2
2^{2(n_1+n_2)}(n_1!n_2!)^2\rho_k^4}}\times \nonumber\\&\exp&\left[\frac{i}{2}\left(\frac{\rho_k^{'}}{\rho_k}+\frac{i}{\rho_k^2}\right)
\left(q_{1{\bf k}}^2+q_{2{\bf k}}^2\right)\right]\nonumber\\
\eea
where $\bar{H}_n$ are the Hermite polynomials of order $n$ and the
associated eigenvalues are given by
\beq\label{eival}
\lambda_{n_1,n_2}=\left(n_1+\frac{1}{2}\right)+\left(n_2+\frac{1}{2}\right)
\eeq

The Lewis phase can be found from its definition
\beq\label{xyz}
\frac{d\alpha_{n_1,n_2}}{d\eta}=\left< \Theta_{n_1,n_2}\left| i
\frac{\partial}{\partial \eta}-\hat{H}_k\right|
\Theta_{n_1,n_2}\right>
\eeq
resulting in \cite{lewis2}
\beq\label{sublewis}
\alpha_{n_1,n_2}=
-\left(n_1+n_2+1\right)\int\frac{d\eta}{\rho^2_k} \eeq
The eigenstates of the  Hamiltonian are now
completely known and are given by
\bea\label{eshamil}
\Psi_{n_1,n_2}&=&\frac{e^{i\alpha_{n_1,n_2}(\eta)}\bar{\textit{H}}_{n_1}\left[\frac{q_{1{\bf
k}}}{\rho_k}\right] \bar{H}_{n_2}\left[\frac{q_{2{\bf
k}}}{\rho_k}\right]}{\sqrt[4]{\pi^2
2^{2(n_1+n_2)}(n_1!n_2!)^2\rho_k^4}}\times \nonumber\\
&\exp&\left[\frac{i}{2}\left(\frac{\rho_k^{'}}{\rho_k}+\frac{i}{\rho_k^2}\right)
\left(q_{1{\bf k}}^2+q_{2{\bf k}}^2\right)\right]\nonumber\\
\eea
Thus one can find the eigenstates of the Hamiltonian for cosmological perturbations for the modes having wavelength
less than the horizon provided it possesses a dynamical invariant containing
no time derivative operation.
\subsection{Modes with $k^2< \frac{z''}{z},\frac{a''}{a}$}
Now to find the solution of the Schroedinger Eqn. \eqref{ham1} for the  modes with $k^2< \frac{z''}{z},\frac{a''}{a}$,
 we shall make use
of the Hamiltonian given by \eqref{suph} with imaginary frequency $i\omega_I$ which is nothing but an inverted harmonic
oscillator which possesses continuous energy spectrum.
The time-dependent inverted harmonic oscillator is exactly solvable just like
the standard time-dependent harmonic oscillator. However, the physics of the time dependent
inverted oscillator is very different \cite{barton, pedro2003, pedro2004}: it has a wholly continuous energy
spectrum varying from minus to plus infinity; its energy eigenstates are no longer
square-integrable and they are doubly degenerate with respect to either the incident
direction or, alternatively, the parity.

 In this case also the invariant operator can be worked out in a similar way as before which turns out to be \cite{pedro2003},
\bea\label{supinvariant}
 I_k&=&\frac{1}{2}\left[
 -\frac{q_{1{\bf k}}^2}{\rho_k^2}+\bigg(\rho_k p_{1{\bf k}}-\rho_k^\prime
 q_{1{\bf k}}\bigg)^2\right]\nonumber\\
 &+&\frac{1}{2}\left[
 -\frac{q_{2{\bf k}}^2}{\rho_k^2}+\bigg(\rho_k p_{2{\bf k}}-\rho_k^\prime
q_{2{\bf k}}\bigg)^2\right]\nonumber\\
 &=& I_{1}+I_{2}
 \eea
where $\rho_k$ now satisfies the following auxiliary equation
\begin{equation}\label{imp}
{\rho_k}^{\prime\prime}-
\omega_I^2(\eta,k)\rho_k=-\frac{1}{\rho^3_k(\eta)}
\end{equation}
Now the eigenstates of the operator $I_k$ is governed by the eigenvalue equation
\bea\label{evet}
I_{\bf k}\Theta_{\lambda_1,\lambda_2}\left(q_{1{\bf k}},q_{2{\bf
k}},\eta\right)=\lambda_{\lambda_1,\lambda_2}\Theta_{\lambda_1,\lambda_2}\left(q_{1{\bf
k}},q_{2{\bf k}},\eta\right)
\eea
Following the steps as in \cite{pedro2003,pedro2004} the eigenstates of the operator $I_k$ turns out to be
\bea\label{supeshamil}
\Theta_{\lambda_1,\lambda_2}&=&\frac{1}{\rho_k}\exp\left[\frac{i}{2}\frac{\rho_k^{'}}{\rho_k}
\left(q_{1{\bf k}}^2+q_{2{\bf k}}^2\right)\right] \nonumber\\
&\times&W_{\lambda_1}\left(\frac{\sqrt{2}q_{1{\bf k}}}{\rho_k},\lambda_1\right)W_{\lambda_2}\left(\frac{\sqrt{2}q_{2{\bf k}}}{\rho_k},\lambda_2\right)\nonumber
\eea

As a result the solution to the Schroedinger equation \eqref{ham1} for the
Hamiltonian \eqref{suph} is now completely known and given by
\bea\label{subeshamil}
\Psi_{\lambda_1,\lambda_2}&=&e^{i\alpha_{\lambda_1,\lambda_2}}\Theta_{\lambda_1,\lambda_2}
\eea
where $W_{\lambda_1}$ and $W_{\lambda_2}$ are parabolic cylinder or Weber functions and
$\alpha_{\lambda_1,\lambda_2}$ are the phase factor i.e. the Lewis phase which are given by
\beq
\alpha_{\lambda_1,\lambda_2}=
-\left(\lambda_1+\lambda_2\right)\int\frac{d\eta}{\rho^2_k}
\eeq

Here we make a point of note that our present framework is not adequate for the calculation of the Berry phase for the super
Hubble modes. Since in this case the system does not possess any well defined ground state. 
 As in the present article 
our primary intention is to show the significance of the geometric phase in the context of inflationary cosmological perturbations
, from now on we restrict our attention to the sub Hubble modes only and try to derive expressions connecting
the Berry phase and cosmological observables.
\section{Berry Phase for the sub Hubble modes}

 To calculate the Berry phase for the sub Hubble modes we shall first make use of the following identity
\beq
\frac{z''}{z} v^2=\left(\frac{z'}{z}\right)^2 v^2-2\frac{z'}{z} vv'+\frac{d}{d\eta}\left[\frac{z'}{z} v^2\right]
\eeq
then the above action \eqref{sa} can be expressed as \cite{pedro}
\beq\label{saas}
{\textbf{ \cal S}}^S=\frac{1}{2}\int d\eta d\textbf{x}\left[v'^2-\delta^{ij}\partial_iv\partial_jv-2\left(\frac{z'}{z}\right)^2vv'+
\frac{z'}{z}v^2\right]
\eeq
which we find more convenient to work with.
Constructing the Hamiltonian we get
\beq
{\bf H}=\frac{1}{2}\int d^3x\left[\Pi^2+\delta^{ij}\partial_iv\partial_jv+2\frac{z'}{z}v\Pi\right]
\eeq
where now  $\Pi= v'-\frac{z'}{z}v$.
Now promoting the fields to operators and taking the Fourier decomposition
we find the Hamiltonian density operator corresponding to the above action \eqref{saas} to be
\bea\label{sho} \hat{\textbf{H}}_{\bf
k}^S&=&\frac{1}{2}\left[\hat{\Pi}_{1\bf
k}^2+\frac{z'}{z}\left(\hat{\Pi}_{1\bf k}\hat{v}_{1\bf
k}+\hat{v}_{1\bf k}\hat{\Pi}_{1\bf k}\right)
+k^2\hat{v}_{1\bf k}^2\right]\nonumber\\
&+&\frac{1}{2}\left[\hat{\Pi}_{2\bf k}^2+\frac{z'}{z}\left(\hat{\Pi}_{2\bf k}\hat{v}_{2\bf k}+\hat{v}_{2\bf k}\hat{\Pi}_{2\bf k}\right)+k^2\hat{v}_{2\bf k}^2\right]\nonumber\\
&\equiv& \hat{\textbf{H}}_{1\bf k}^S+\hat{\textbf{H}}_{2\bf k}^S
\eea 

Similarly the Hamiltonian operator corresponding to the tensor perturbations is found to be
\bea\label{tho}
\hat{\textbf{H}}_{\bf k}^T&=&\frac{1}{2}\left[\hat{\pi}_{1\bf k}^2+\frac{a'}{a}\bigg(\hat{\pi}_{1\bf k}\hat{u}_{1\bf k}+\hat{u}_{1\bf k}\hat{\pi}_{1\bf k}\bigg)+k^2\hat{u}_{1\bf k}^2\right]\nonumber\\
&+&\frac{1}{2}\left[\hat{\pi}_{2\bf k}^2+\frac{a'}{a}\bigg(\hat{\pi}_{2\bf k}\hat{u}_{2\bf k}+\hat{u}_{2\bf k}\hat{\pi}_{2\bf k}\bigg)+k^2\hat{u}_{2\bf k}^2\right]\nonumber\\
&\equiv& \hat{\textbf{H}}_{1\bf k}^T+\hat{\textbf{H}}_{2\bf k}^T
\eea

In a compact general form the Hamiltonians can be written as a sum of two generalized time dependent harmonic oscillators as
\beq\label{ho} \hat{\textbf{H}}_{j\bf
k}=\frac{1}{2}\bigg[k^2\hat{q}_{j\bf
k}^2+Y(\eta)\left(\hat{p}_{j\bf k}\hat{q}_{j\bf k}+\hat{q}_{j\bf
k}\hat{p}_{j\bf k}\right)+\hat{p}_{j\bf k}^2\bigg]
\eeq
where
$\hat{q}_{j\bf k}=\hat{v}_{j\bf k}$, $\hat{u}_{j\bf k}$;
$\hat{p}_{j\bf k}=\hat{\Pi}_{j\bf k}$, $\hat{\pi}_{j\bf k}$  and
$Y=\frac{z'}{z}$,  $\frac{a'}{a}$ for the scalar and tensor modes
respectively and $j=1,2$ with the frequency given by
$\omega=\sqrt{k^2-Y^2}$.

Following the same trail \cite{lewis1,lewis2,lewis3} we find
\bea\label{invariant}
 I_k&=&\frac{1}{2}\left[
 \frac{q_{1{\bf k}}^2}{\rho_k^2}+\bigg(\rho_k \bigg[p_{1{\bf k}}+Yq_{1{\bf k}}\bigg]-\rho_k^\prime
 q_{1{\bf k}}\bigg)^2\right]\nonumber\\
 &+&\frac{1}{2}\left[
 \frac{q_{2{\bf k}}^2}{\rho_k^2}+\bigg(\rho_k \bigg[p_{2{\bf k}}+Yq_{2{\bf k}}\bigg]-\rho_k^\prime
q_{2{\bf k}}\bigg)^2\right]\nonumber\\
 &=& I_{1}+I_{2}
 \eea
where $\rho_k$ now satisfies the following equation
\begin{equation}\label{submp}
{\rho_k}^{\prime\prime}+
\Omega^2(\eta,k)\rho_k=\frac{1}{\rho^3_k(\eta)}
\end{equation}
with $\Omega^2=\omega^2-\frac{dY}{d\eta}$.

The eigenstates of the operator $I_k$ turn out to be
\bea\label{subtes}
\Theta_{n_1,n_2}&=&\frac{\bar{H}_{n_1}\left[\frac{q_{1{\bf
k}}}{\rho_k}\right]\bar{H}_{n_2}\left[\frac{q_{2{\bf
k}}}{\rho_k}\right]}{\sqrt[4]{\pi^2
2^{2(n_1+n_2)}(n_1!n_2!)^2\rho_k^4}}\times \nonumber\\&\exp&\left[\frac{i}{2}\left(\frac{\rho_k^{'}}{\rho_k}-Y(\eta)+\frac{i}{\rho_k^2}\right)
\left(q_{1{\bf k}}^2+q_{2{\bf k}}^2\right)\right]\nonumber\\
\eea

As a consequence the eigenstates of the  Hamiltonian are now given by
\bea\label{subeshamil}
\Psi_{n_1,n_2}&=&e^{i\alpha_{n_1,n_2}(\eta)}\Theta_{n_1,n_2 }
\eea
where the Lewis phases are given by \eqref{sublewis}. The phase $\alpha_{n_1,n_2}(\eta)$ is the combination of the
dynamical phase and the geometric phase which can be well understood from the Eqn.\eqref{xyz}. Once the Lewis phase is calculated, this can be utilized in deriving the geometric phase associated with the system corresponding to the
particle creation through the vacuum quantum fluctuations during inflation.

But before proceeding in this direction, we would like to present the general wavefunction for the
vacuum state of the inflationary cosmological perturbations. To this end, let us consider the parametric harmonic
oscillator Hamiltonian
\beq\label{sho1}
\hat{\textbf{H}}_{\bf
k}^S=\frac{1}{2}\left[\hat{\Pi}_{\bf
k}^2+\frac{z'}{z}\left(\hat{\Pi}_{\bf k}\hat{v}_{\bf
k}+\hat{v}_{\bf k}\hat{\Pi}_{\bf k}\right)
+k^2\hat{v}_{\bf k}^2\right]
\eeq
which refers to
\eqref{sho} and can be solved analytically for the vacuum. By the following similarity transformation
\beq
\hat{\tilde{H}}_{\bf k}\equiv A\tilde{H}_{\bf k}A^{-1},~ \mbox{where}~~ A=e^{-i\frac{z^\prime}{2z} v_{\bf k}^2}
\eeq
the Hamiltonian can be reduced to the following form
\beq\label{sho2}
\tilde{\textbf{H}}_{\bf
k}^S=\frac{1}{2}\left[\hat{\Pi}_{\bf
k}^2+(k^2-\frac{z'^2}{z^2})\hat{v}_{\bf k}^2\right]
\eeq

The wavefunction for the Hamiltonian \eqref{sho2} is quite well known and has the form \cite{martincos}
\beq
{\tilde{\Psi}}_k=N_k e^{-\Omega_kv_k^2}
\eeq
where
\beq
|N_k|=\left( \frac{2Re\Omega_k}{\pi}\right)^{1/4}, ~~~\Omega_k=-\frac{i}{2}\frac{f_k^\prime}{f_k}
\eeq
and $f_k$ satisfies
\beq
f_k^{\prime\prime}+\left(k^2-\frac{z'^2}{z^2}\right)f_k=0
\eeq
For the vacuum state we know
\beq
f_k=\frac{1}{\sqrt{2k}}e^{ik\eta}
\eeq
which gives us
\beq
{\tilde{\Psi}}_k=\left( \frac{k}{\pi}\right)^{1/4}e^{-\frac{k}{2}v_k^2}.
\eeq
Hence  the vacuum state wavefunction for the inflationary cosmological scalar perturbations turns out to be
\beq
\Psi_k^S=\left( \frac{k}{\pi}\right)^{1/2}e^{-(i\frac{z^\prime}{z}+k)(v_{1k}^2+v_{2k}^2)}
\eeq
Similarly, we can write the vacuum state wavefunction for the inflationary cosmological tensor  perturbation as
\beq
\Psi_k^T=\left( \frac{k}{\pi}\right)^{1/2}e^{-(i\frac{a^\prime}{a}+k)(v_{1k}^2+v_{2k}^2)}
\eeq
 The phase part, which is  the combination of the dynamical phase and the geometric phase, of the wavefunction is now explicit but it is not easy to separate out the geometric phase from this expression.
But in our present framework this can be done using the expressions already derived, which is what we do next.




 With the help of  Eqns.(\ref{subtes}) and \eqref{sublewis} we obtain the corresponding Berry
phase
\bea\label{gp}
\gamma_{n_1,n_2,k}&\equiv&i\int_0^\Gamma\left<
\Theta_{n_1,n_2}\left|\frac{\partial}{\partial \eta}\right|
\Theta_{n_1,n_2}\right> d\eta\\
&=&-\frac{1}{2}(n_1+n_2+1)\int_0^\Gamma
\left(\frac{1}{\rho_k^2}-\rho_k^2\omega^2-(\rho_k^\prime)^2\right)d\eta\nonumber
\eea
 where it has been assumed that the invariant $I_k(\eta)$ is
$\Gamma$ periodic and its eigenvalues are non-degenerate.

To get a
deeper physical insight the quantitative estimation of the Berry
phase is very important.
 Eqn.(\ref{gp}) tells us that for this estimation, the knowledge of
$\rho_k$ is essential but the solution of Eqn.\eqref{mp} is difficult to obtain.
Another point to be carefully handled is to set the value of the parameter $\Gamma$.
Keeping all these in mind and considering compatible physical conditions we
proceed as follows.

First we note that in the adiabatic limit (which is quite justified for sub-Hubble modes) Eqn.\eqref{mp} can be
solved \cite{lewis2} by a series of powers in adiabatic parameter,
$\delta$ ($<<1$). To this end we define a slowly varying time variable as
$\tau=\delta\eta$ and write the solution to Milne-Pinney equation as
\beq
\rho_k=\rho_0+\delta\rho_1+\delta^2\rho_2+...
\eeq
Inserting this expansion into the Milne-Pinney equation we get
\bea\label{mpsol}
\delta^2\rho_0\ddot{\rho_0}&+&\rho_0^2\left[1+2\delta\rho_0\rho_1+\delta^2\rho_1^2+2\delta^2\rho_0\rho_2\right]\left(\omega^2-\delta \dot{Y}\right)\nonumber\\
&=&\frac{1}{\rho_0^2+2\delta\rho_0\rho_1+\delta^2\rho_1^2+2\delta^2\rho_0\rho_2}+O(\delta^3)
\eea
Here dot represents derivative w.r.t. new time variable $\tau$.
Collecting the zeroth order terms from both sides we obtain $\rho_0^2=\omega^{-\frac{1}{2}}$.
Now the integrand of Eqn.\ref{gp} can be rewritten as
\bea
\frac{1}{\rho_k^2}&-&\rho_k^2\omega^2-(\rho_k^\prime)^2=\rho_k\rho_k''-(\rho_k')^2-\rho_k^2Y'\nonumber\\
&\simeq&\delta \rho_0^2\dot{Y}+\delta^2\left[\rho_0\ddot{\rho_0}-\dot{\rho}^2-2\rho_0\rho_1\dot{Y}\right]+O(\delta^3)\nonumber\\
&=&\delta\omega^{-\frac{1}{2}}\dot{Y}+O(\delta^2)
\eea

Thus for the ground state of the system, in the adiabatic limit, the Berry phase for a particular perturbation mode can be evaluated
 upto the first order in $\delta$, which is given by
\begin{equation}\label{bp}
\gamma_{k}^{(S, ~T)}=-\frac{1}{2}\int_0^\Gamma
\frac{\delta\dot{Y}}{\sqrt{k^2-Y^2}}d\eta=-\frac{1}{2}\int_0^\Gamma
\frac{Y'}{\sqrt{k^2-Y^2}}d\eta
\end{equation}
where the superscripts $S$ and $T$ stand for scalar and tensor
modes respectively.
One may note that our result \eqref{bp} coincides with that of Berry \cite{berry1}.

Our next task is to fix the value of the parameter $\Gamma$. To this end we
shall calculate the total Berry phase accumulated by each mode during sub Hubble evolution in the inflationary era. For
the ground state of the system this turns out to be
\beq\label{bpacc1}
\gamma_{k~sub}^{S,T}=-\frac{1}{2}\lim_{\eta'\rightarrow
-\infty}\int_{\eta'}^{\eta_0^{S,T}}
\frac{Y'}{\sqrt{k^2-Y^2}}d\eta
\eeq
where $\eta_0^{S,T}$ is the conformal time which satisfies the relation $k^2=\left[Y(\eta_0^{S,T})\right]^2$
so that the modes are within the horizon and oscillating with real frequencies.
A non-zero value of the parameter $\gamma_{k~sub}^{S,T}$ will ensure that there are some nontrivial effects
of the curved space-time background on the evolution of the quantum fluctuations and may play an
important role in the growth of inflationary cosmological perturbations.

\section{Berry phase and the cosmological parameters}

Let us now set up the link between this cosmological analogue of Berry phase
and the cosmological observables.
In the adiabatic limit, {\it accumulated Berry phase} during sub-Hubble oscillations of the each mode
is given by \eqref{bpacc1}.
The formula \eqref{bpacc1} is adopted to derive the relations between the
accumulated Berry phase and the cosmological observable parameters. From now on
we shall drop the subscript `$sub$' keeping in mind that the calculations are for sub-horizon modes only.

Now the variable $Y(\eta)$ can be expressed in-terms of
the slow-roll parameters. If we neglect the time variation in slow-roll parameters then Eqn.\eqref{bpacc1} can be
integrated analytically.
Then the {\it accumulated Berry phase} during sub-Hubble evolution of the scalar modes, in terms of the slow-roll
parameters, turns out to be
\bea
\gamma_k^S &=& \frac{1}{2}\lim_{\eta'\rightarrow-\infty}\int_{\eta'}^{-\frac{\sqrt{1+6\epsilon_1-2\epsilon_2}}{k}}
\frac{\frac{z^{''}}{z}- \left(\frac{z^{'}}{z} \right)^{2}}{\sqrt{k^2-\left(\frac{z^{'}}{z} \right)^{2}}}d\eta \nonumber\\
&=& \frac{1}{2}\lim_{\eta'\rightarrow-\infty}\int_{\eta'}^{-\frac{\sqrt{1+6\epsilon_1-2\epsilon_2}}{k}}\frac{\eta^{-2}\left(1+3\epsilon_1-\epsilon_2\right)}
{\sqrt{k^2-\eta^{-2}\left(1+6\epsilon_1-2\epsilon_2\right)}}d\eta \nonumber\\&+&O(\epsilon_1^2,\epsilon_2^2,\epsilon_1\epsilon_2)\nonumber\\
&\approx&
-\frac{\pi}{4}\frac{1+3\epsilon_1-\epsilon_2}{\sqrt{1+6\epsilon_1-2\epsilon_2}}\label{scal}
\eea
For brevity, we have restricted our analysis upto the first order in
{\it slow-roll} parameters and we have neglected any time variation in
 $\epsilon_1,~\epsilon_2$. And for the tensor modes we have
\bea\label{tens}
\gamma_k^T &=& \frac{1}{2}\lim_{\eta'\rightarrow-\infty}\int_{\eta'}^{-\frac{\sqrt{1+2\epsilon_1}}{k}} \frac{\frac{a^{''}}{a}- \left(\frac{a^{'}}{a} \right)^{2}}{\sqrt{k^2-\left(\frac{a^{'}}{a} \right)^{2}}}d\eta\nonumber\\
&=&\frac{1}{2}\lim_{\eta'\rightarrow-\infty}\int_{\eta'}^{-\frac{\sqrt{1+2\epsilon_1}}{k}} \frac{\eta^{-2}\left(1+\epsilon_1\right)}{\sqrt{k^2-\eta^{-2}\left(1+2\epsilon_1\right)}}d\eta\nonumber\\
&+&O(\epsilon_1^2,\epsilon_2^2,\epsilon_1\epsilon_2)\nonumber\\
&\approx& -\frac{\pi}{4}\frac{1+\epsilon_1}{\sqrt{1+2\epsilon_1}}
\eea
Here also we have neglected any time variation in
 $\epsilon_1,~\epsilon_2$ and restricted our attention upto the first order in them.
In the above derivations we have made use of the standard definition
of the slow-roll parameters \cite{lb}
\bea
\epsilon_1\equiv\frac{M_P^2}{2}\left(\frac{V^{'}(\phi)}{V(\phi)}\right)^2,
~~ \epsilon_2\equiv M_P^2\left(\frac{V^{''}(\phi)}{V(\phi)}\right),
\eea
 $V(\phi)$ being inflaton potential.
For the estimation of $\gamma^{S,T}_k$, the slow-roll parameters are
to be evaluated at the start of inflation. But during inflation the
slow-roll parameters does not evolve significantly from their
initial values for first few $e$-folds, which is relevant for the
present day observable modes as they are supposed to leave the horizon during first 10 e-folds. So in the above estimates for
$\gamma^{S,T}_k$ we can consider $\epsilon_1$ and $\epsilon_2$
as their values at horizon crossing without committing any substantial error.

We are now in a position to relate this phase with observable parameters.
At the horizon exit the fundamental observable parameters can be expressed in terms of
slow-roll parameters (upto the first order in
$\epsilon_1,~\epsilon_2$) as \cite{stewart,lb,andrew}
\bea
P_{\cal R}&=&\frac{V}{24\pi^2M^4 \epsilon_1},~~n_S=1+2\epsilon_2-6\epsilon_1\nonumber\\
n_T&=&-2\epsilon_1,~~r=16\epsilon_1 \label{obssr}
\eea
where $P_{\cal R}$ is the scalar power spectrum, $n_S$ and $n_T$ are the scalar and tensor spectral indices respectively, $r$ is the tensor to scalar ratio.
As a consequence the {\it accumulated Berry phase} associated with the {\it sub-Hubble} oscillations of the scalar fluctuations during inflation
can be expressed in terms of the observable parameters (and vice versa) using \eqref{scal} and \eqref{obssr}  as follows
\bea\label{bpt1}
\gamma^S_{k}
&\approx&-\frac{\pi}{8}\frac{3-n_S(k)}{\sqrt{2-n_S(k)}}\\
n_S(k)&\approx&3-\frac{8\gamma^S_{k}}{\pi}\left(\frac{4\gamma^S_{k}}{\pi}-\sqrt{\frac{16[\gamma^S_{k}]^2}{\pi^2}-1}\right)
\eea
Therefore accumulated Berry phase for the scalar modes is
related to the scalar spectral index. From the above relation it is also very clear
how the Berry phase is related to the cosmological curvature perturbations. For the tensor modes using \eqref{tens} and \eqref{obssr} the corresponding expressions turn
out to be
\bea\label{bpt2}
\gamma^T_{k}
&\approx&-\frac{\pi}{8}\frac{2-n_T(k)}{\sqrt{1-n_T(k)}}\\
n_T(k)&\approx& 2-\frac{8\gamma^T_{k}}{\pi}\left(\frac{4\gamma^T_{k}}{\pi}-\sqrt{\frac{16[\gamma^T_{k}]^2}{\pi^2}-1}\right)
\eea
Eqns.\eqref{bpt1} and \eqref{bpt2} reveal that the Berry phase due to
scalar and tensor modes basically correspond to a new parameter made of
corresponding spectral indices. Here we note that the relations \eqref{bpt1}, \eqref{bpt2} are not exact in general
but they are in the linearized theory of cosmological perturbation. Had we taken into
account the second and higher order contributions of the cosmological fluctuations the relations \eqref{bpt1}, \eqref{bpt2}
would have been different.

Further, the accumulated Berry phase associated with the total gravitational fluctuations
(a sum-total of  $\gamma_{k}^S$ and $\gamma_{k}^T$) can be expressed in terms of
the other observable parameter as well, giving
\bea
\gamma_{k}&\equiv&\gamma_{k}^S+\gamma_{k}^T\approx-\frac{\pi}{8}\left[\frac{3-n_S(k)}{\sqrt{2-n_S(k)}}+\frac{2+\frac{r}{8}}{\sqrt{1+\frac{r}{8}}}\right] \label{bpt3} \\
&\approx&-\frac{\pi}{8}\left[\frac{3-n_S(k)}{\sqrt{2-n_S(k)}}+\frac{2+\frac{V}{12\pi^2M_P^4 P_{\cal R}}}{\sqrt{1+\frac{V}{12\pi^2M_P^4 P_{\cal R}}}}\right] \label{bpt4}
\eea
Therefore the {\it accumulated Berry
phase} for sub-Hubble oscillations of the perturbation modes during
inflation can be completely envisioned through the observable
parameters. Here also we see that the total Berry phase of a single mode can be
characterized by the curvature perturbations. The estimation of the Berry phase gives a deeper physical
insight 
of the quantum property of the inflationary perturbation modes. 
As a result, at least in principle, we can claim that measurement of Berry phase can serve as a probe of quantum properties reflected on classical observables.

\section{Physical Significance of cosmological Berry phase}
The physical implication of the Berry phase in cosmology is already transparent from our above analysis.
In a nutshell, the classical cosmological perturbation modes (both scalar and tensor) having quantum origin picks
up a phase during their advancement through the curved space-time background that depends entirely on the background
geometry and may be, at least in principle, estimated quantitatively by measuring the corresponding spectral indices.
So the Berry phase for the quantum counterpart of the classical cosmological perturbations endows us with the measure
of spectral index.

Also, the existing literature suggests that there may be an intriguing  direct link of the cosmological Berry phase
with the CMB. The interpretation of the Wigner rotation matrix as the Berry phase \cite{mukund} has already been
elaborated by the proposal of an optical demonstration \cite{manzon}. On the other-hand, Wigner rotation matrix
can be represented as a measure of statistical isotropy violation of the temperature fluctuations in CMB \cite{tarun}. These results
motivate us to investigate  whether the cosmological analogue of the Berry phase may be thought of as a measure of
violation of statistical isotropy in CMB as our future project.


The current observations from WMAP7 \cite{wmap} have put stringent
constraints on $n_S$ $(0.948 < n_S <1)$ but only an upper bound for
$r$ has been reported so far ($r < 0.36$ at $95 \%$ C.L.), with
PLANCK \cite{planck} expecting to  survey upto the order of
$10^{-2}$. Given this status, any attempt towards the measurement of cosmological Berry
phase may thus reflect observational credentials of this parameter
in inflationary cosmology. For example, it is now well-known that
any conclusive comment on the energy scale of inflation ($V$ in Eqn.
(\ref{bpt3})) provides crucial information about fundamental
physics. However, in CMB polarization experiments, the energy scale
cannot be conclusively determined because there is a degeneracy
between E and B modes via the first slow roll parameter $\epsilon_1$
(Eqn. \ref{obssr}), which can only be sorted out once $r$ is
measured conclusively. But B mode polarized states can be
contaminated with cosmic strings, primordial magnetic field etc,
thereby making it difficult to measure $r$ conclusively (for a lucid
discussion see \cite{challinor}).
So, cosmological Berry
phase may have the potentiality to play some important role in
inflationary cosmology, since it is related to $r$ and $V$ via Eqns. \eqref{bpt3} and \eqref{bpt4}.


\section{Conclusion  }
In this article we have demonstrated how the exact expression for the
wave function of the quantum cosmological perturbations can be
analytically obtained by solving the associated Schroedinger
equation following the dynamical invariant technique. This helps us
to derive an expression for cosmological analogue of Berry phase.
Finally, we demonstrate  how this quantity is related to cosmological parameters and show the physical significance of the cosmological Berry phase.

So far as the detection of cosmological Berry phase is concerned, we
are far away from quantitative measurements. A possible theoretical
aspect of detection \cite{squeeze} of the analogue of cosmological
Berry phase may be developed in squeezed state formalism
\cite{pedro}. On principle Berry phase can be measured  from an experiment dealing
with phase difference (e.g. interference).  Recently, an analogy between phonons in an
axially time-dependent ion trap and quantum fields in an
expanding/contracting universe has been derived and corresponding
detection scheme for the analogue of cosmological particle creation
has been proposed which is feasible with present-day technology
\cite{schultz}. Besides, there exists \cite{pedro} a scheme for
measuring the Berry phase in the vibrational degree of freedom of a
trapped ion. We hope that these type of detection schemes may be
helpful for observation of the cosmological analogue of the Berry
phase in laboratory in future.

It is hoped that further research in this direction may help us to anticipate how these relations can be utilized in extracting further information related to the theoretical and observational aspects of inflationary perturbations.


\section*{Acknowledgments}
BKP thanks Council of Scientific and Industrial Research,
India for financial support through Senior Research Fellowship
(Grant No. 09/093 (0119)/2009). Part of SP's work is supported by a
research grant from Alexander von Humboldt Foundation, Germany,
and by the SFB-Tansregio TR33 ``The Dark
Universe'' (Deutsche Forschungsgemeinschaft) and the European
Union 7th network program ``Unification in the LHC era''
(PITN-GA-2009-237920). 

\end{document}